
\input phyzzx
\def\rl{\rightline}
\tolerance=10000
\singlespace
\rl{WIS--93/21/MAR--PH}
\rl{June 20, 1993}
\rl{T}
\pagenumber=0
\normalspace
\bigskip
\titlestyle{\bf TECHNIDILATON PHENOMENOLOGY AND PROSPECTS FOR PRODUCTION}
\vskip 18pt
\author{Edi Halyo {\footnote{@}{e-mail address: jphalyo@weizmann.bitnet.}}}
\smallskip
\centerline{\it Department of Physics, Weizmann Institute of Science}
\centerline{\it Rehovot 76100, ISRAEL}
\vskip 5 truecm
\titlestyle{\bf ABSTRACT}
An effective Lagrangian for the technidilaton and its interactions with
matter is constructed. Properties of the technidilaton are compared with
those of the Higgs boson. Technidilaton decays and production channels are
investigated. Main technidilaton decays are suppressed compared to the Higgs
boson and the most important production mechanism is due to gluon
fusion. Prospects for technidilaton production and detection at $e^+e^-$
colliders and the SSC are examined. LEP 1, LEP 2 and SSC can find or rule out
a technidilaton with a mass up to $10~GeV$, $40~GeV$ and $230~GeV$
respectively.

\bigskip

\vfill
\eject

\centerline{\bf 1. Introduction}

It is generally believed that the observation of a low--mass scalar particle
will confirm the existence of a Higgs boson with or without supersymmetry. In
particular it is argued that this will refute technicolor (TC) [1] theories
which have no fundemental scalars. In this paper we consider an alternative
possibility in which a low--mass scalar exists but it is not the Higgs boson.

In previous papers [2,3] scale symmetry properties of TC were
investigated. It was found that due to the spontaneous and anomalous scale
symmetry breaking there is a pseudo--Goldstone boson of dilatations, a
technidilaton, in TC. The technidilaton mass was estimated by using Dashen's
theorem and its interactions were obtained by constructing a low--energy
effective
Lagrangian. The technidilaton's interactions were found to be very similar to
those of the Higgs boson with minor but important differences.

In this paper we investigate the decay and production mechanisms of the
technidilaton and compare them with those of the Higgs boson. We find that all
main technidilaton decay rates are suppressed by a factor of $v^2/F_D^2 \sim
0.3/N$, for a $SU(N)_{TC}$ group, relative those of the Higgs. Therefore the
technidilaton is much narrower (unless it can decay into light pseudo--
Goldstone bosons) and more difficult to detect. Because of the
same suppression factor, it is much harder to produce technidilatons compared
to Higgs bosons for a given mass at a given energy. We find that, for the
technidilaton, the only
production mechanism not supressed is that by gluon fusion. This is a
consequence of the large, anomalous technidilaton coupling to two gluons which
arises from the trace anomaly in QCD. As a result, $e^+e^-$ colliders are not
efficient for finding or ruling out the technidilaton. A $ p p$ collider like
the SSC, on the other hand, is more efficient
since the main production mechanism which is gluon fusion is not suppressed.
We discuss prospects for technidilaton production at $e^+e^-$ colliders and
at the SSC. We find that LEP 1 and LEP 2 can only find or rule out a
technidilaton
with a mass of $10~GeV$ and $40~GeV$ respectively. At the SSC, a wider and
higher mass range can be searched. Contrary to $e^+e^-$ colliders, the results
for SSC are model dependent for two reasons. First, at the relevant
technidilaton masses, decays into light axions of the TC model are important.
The number of these light axions and their masses depend on the specific TC
model considered. Second, the QCD $\beta$--function and therefore the
technidilaton coupling to two gluons due to the trace anomaly
is model dependent. We consider two generic TC models
and find that SSC can find a technidilaton with a mass up to
$\sim 230~GeV$. In the first model this whole range can be searched whereas in
the second one the window $100~GeV<M_D<2M_W$ remains.

The paper is organized as follows. Section 2 is a review of the technidilaton's
properties and interactions. In Section 3, we obtain the
different decay rates and production cross sections for the technidilaton.
These are compared with those for the Higgs boson. In Section 4 we discuss
prospects for technidilaton production and observation at present and future
colliders such as
LEP 1, LEP 2, TLC and SSC. Section 5 is a discussion of our results.

\bigskip
\centerline{\bf  2. Technidilaton and its interactions}

For simplicity consider the one-doublet TC model [4] with an $SU(N)$ gauge
group and two massless technifermions $A^i, B^i, i=1,\ldots ,N$ where $i$ is
the $SU(N)_{TC}$ index. Everything in the following discussion applies equally
well to the cases with more than one doublet but then the notation gets very
cumbersome without any new insight gained. The two technifermions are color
singlets which form an $SU(2)_L$ doublet $T_L=\left( \matrix{A^i \cr B^i \cr}
\right)_L$ and two $SU(2)_L$ singlets $A^i_R, B^i_R$.
This theory is an exact analog of ordinary $QCD$
with one doublet of quarks where the gauge group is $SU(N)$ instead of $SU(3)$
and all energies are scaled up by a factor of $\simeq 2600$. At energies around
$\Lambda_{TC} \simeq 1~TeV $, TC gets strong and confining; as a result
only TC singlets appear at lower energies. In addition, a fermion condensate
 $\langle 0|\bar{A}^i A^i+
\bar{B}^i B^i|0\rangle$ with a non-zero vacuum expectation value which
breaks the gauge symmetry from $SU(2)_L \times U(1)_Y$
to $U(1)_Q$ forms. This condensate also breaks the global symmetry
from $SU(2)_L \times
SU(2)_R$ to $SU(2)_V$ causing three technipions $\Pi^+=\bar{A}^i A^i,\quad
\Pi^-=\bar{B}^i B^i,\quad  \Pi^0={1\over \sqrt2}(\bar{A}^i A^i+
\bar{B}^i B^i) $ to form. The technipions are eaten by the gauge bosons and
become their longitudinal components. At low energies ({\it i.e.} $E<M_W$),
the technipions behave as the longitudinal components of the gauge bosons.
But at high energies ({\it i.e.} $E>>M_W$) they behave as technipions: they
are pseudoscalars which interact srongly among themselves. There are other
TC singlets [4] ({\it e.g.} technihadrons, vector technimesons) but since they
are not Goldstone bosons they have larger masses ($\sim 1.5-2~ TeV$). As a
result at energies which are low compared to the TC scale
($E<<\Lambda_{TC}$) TC is effectively described by the technipions and
their interactions.

The scale symmetry properties of TC were
investigated in Ref. [2]. The results obtained can be summarized
as follows. Scale symmetry is
broken spontaneously and anomalously in TC. The spontaneous symmetry breaking
is due to the technifermion vacuum condensate $\langle 0|\bar{A}^i A^i+
\bar{B}^i B^i|0\rangle$ and the technigluon vacuum condensate
$\langle 0 | F^{\mu\nu a}_{\rm TC} F^a_{\mu\nu\,{\rm TC}} |0\rangle$.
The technigluon condensate dominates spontaneous scale symmetry
breaking due to the large Casimir $C_2$ in the two technigluon channel.
In addition, scale symmetry is broken anomalously i.e. there is an
anomaly in the divergence of the scale current $\partial_\mu D^\mu$ of the
theory which is given by [5]
$$\partial_\mu D^\mu = \Theta^\mu {}_\mu =
{\beta(\alpha_{TC}) \over {4\alpha_{TC}}} F^{\mu\nu a}_{TC} F^a_{\mu\nu TC}
\qquad a=1,\ldots,N^2-1 \eqno(1)$$
where $\Theta^\mu {}_\mu$ is the trace of the new improved energy-momentum
tensor [6], $\alpha_{TC}$ and $\beta(\alpha_{TC})$ are the TC coupling constant
and the $\beta$-function respectively. $F^{\mu\nu}_{TC}$ is the TC field
strength tensor. Since the coefficient of the scale
anomaly in Eq. (1) is much smaller than one at the spontaneous symmetry
breaking
scale (which is given by $C_2(FF)\alpha_{TC}(Q^2=\Lambda_D^2) \sim 1$) one can
talk about a spontaneously broken approximate scale symmetry.
As a result, there is a pseudo-Goldstone boson of dilatations, a
technidilaton, in TC. The technidilaton acquires a small mass because of the
small explicit scale symmetry breaking due to the scale anomaly. The mass of
the technidilaton can be estimated by
using PCDC (partially conserved dilatation current hypothesis) and Dashen's
theorem [7] $$M_D^2= -{1\over F_D^2} \langle 0| {\beta(\alpha_{TC})
\over \alpha_{TC}} F^{\mu\nu a}_{TC} F^a_{\mu\nu TC} |0\rangle \eqno(2) $$
where $F_D$ is the technidilaton constant.
The numerical value for $M_D$ can be estimated by scaling the results of the
QCD case and using the $1/N$ expansion [8]. One gets $M_D \simeq (7.5/Nr)~TeV$
for an $SU(N)_{TC}$ gauge group with $r$ technifermion doublets.
Therefore, in
addition to the usual light pseudoscalar pseudo-Goldstone bosons (PGBs) there
is
a light scalar boson in the low--energy spectrum of TC.

In Ref. [3] the interactions of the technidilaton with matter were
obtained by constructing a low--energy ({\it i.e.} $E< 1~TeV$ ) effective
Lagrangian which realizes both chiral and scale symmetries non-linearly.
Methods for constructing an effective Lagrangian which realizes spontaneously
broken chiral symmetries and scale symmetry are well known [9,10].

The effective technipion Lagrangian is analogous to the chiral Lagrangian [9]
which is the effective low-energy Lagrangian for the ordinary pions in QCD.
The only differences are the replacement of the pion constant $f_\pi$ by
the technipion constant $F_\pi \simeq 246~ GeV$ and the fact that the theory
is gauged to $SU(2)_L \times U(1)_Y$.
The effective Lagrangian is given by
$${\cal L}_{eff}={F_\pi^2 \over 4} \Tr \left[(D_\mu \Sigma)(D^\mu \Sigma)
^\dagger \right]  \eqno (3) $$ where $\Sigma=e^{2\tilde\Pi i / F_\pi}$
and $$\tilde\Pi= \Pi^a T^a={1\over \sqrt 2} \left( \matrix {{\Pi^0\over \sqrt2}
& \Pi^+ \cr\noalign{\vskip 0.2 cm} \Pi^- & {\Pi^0\over \sqrt2} \cr}\right)
\eqno (4)$$ are the technipions. The covariant derivative is
$D_\mu= \partial_\mu - igW^a_\mu T^a -ig'YB_\mu$ since under $SU(2)_L$
$\Sigma \to \Sigma'=\Sigma+i\alpha^a_L T^a \Sigma $ and under $U(1)_Y$
$\Sigma \to \Sigma'=\Sigma+i \alpha_Y Y \Sigma $
where $W^a_\mu$ and $B_\mu$
are the $SU(2)_L$ and $U(1)_Y$ gauge fields, $\alpha^a_L$ and $\alpha_Y$
are the parameters of the two transformations. ${\cal L}_{eff}$ is valid for
the energy range $M_W << E << 4\pi F_\pi$ ({\it i.e.} $ 80~ GeV << E <<3~
TeV$).
The lower limit arises from the fact that it is only for $E >>M_W$ that the
longitudinal components of the gauge bosons behave as technipions. The upper
limit is a result of the fact that only for $E <<4\pi F_\pi$ higher order terms
in the expansion of ${\cal L}_{eff}$ can be neglected with respect to lower
order terms.
The unitary gauge can easily be
employed by taking $\Sigma=1$ which simply means that the Goldstone bosons
({\it i.e.} the technipions) disappear from the spectrum. In unitary gauge
${\cal L}_{eff}$ given by Eq. (3) reduces to the weak gauge boson mass terms.

In order to obtain the interactions of the technidilaton with technipions
({\it i.e.\/} longitudinal gauge bosons) and transverse gauge bosons one
constructs an effective Lagrangian for the energy range
$E<\!\!<\hbox{min}\,\left[ 4\pi F_\pi, 4\pi F_D\right]$.  This effective
Lagrangian must realize both spontaneously broken chiral and scalar symmetries
non-linearly.

Consider a scalar field $U= e^{D/F_D}$ with scale dimension one [10].
$D$ is the
technidilaton field and $F_D$ is the technidilaton constant.  Under
infinitesimal dilatations $\delta_D x^\mu = - x^\mu $,
$U$ transforms as a field with scale dimension one
$\delta_D U = \left( 1 + x^\mu \partial_\mu\right)U$
which gives the technidilaton transformation
$\delta_D D = F_D + \left( x^\mu \partial_\mu\right) D$.
The technidilaton Lagrangian is given by
$${\cal L}_D = {F^2_D\over 2}\partial_\mu U \partial^\mu U - {F^2_DM^2_D\over
4} U^2 \eqno(5)$$
where the first term is the scale invariant kinetic term and the second one
is the technidilaton mass term which breaks dilatations.

The technipions which realize the spontaneously broken $SU(2)_A$
non-linearly have scale dimension zero [11] which means that their kinetic
term,
given by Eq. (3) is not scale invariant.  But above the TC scale or for $E>4\pi
F_\pi$, TC is a theory of massless fermions and gauge bosons which is
classically scale invariant.  In particular the fermion and gauge boson
kinetic terms are scale
invariant.  Therefore the technipion kinetic term given by Eq. (3) is
rendered scale invariant by multiplying by $U^2$ [10]. Then
$${\cal L}_{\rm kin} = {F^2_\pi\over 4} \Tr\left[ \left( D_\mu\Sigma\right)
\left( D^\mu\Sigma\right)^\dagger \right] U^2 \ \ .\eqno(6)$$
This term gives the technipion-technidilaton and gauge boson-technidilaton
interactions. For example, in the
unitary gauge defined by $\Sigma =1$ where the technipions are ``eaten''
$$\eqalign{{\cal L}_{\rm kin} &= {\cal L}_{\rm mass} U^2 \cr
&= \left( M^2_W W^+_\mu W^{-\mu} + {1\over 2} M^2_Z Z^{02} \right) \left( 1 +
{2D\over F_D} + {2D^2\over F^2_D}+\ldots\right)\ \ \cr}  \eqno(7)$$
The one and two technidilaton couplings to gauge bosons can be read
immediately from the expression above.

If there are more than one technifermion doublet, then not all Goldstone
bosons are ``eaten''.  For example, in the one-family model [4] there are
63 Goldstone bosons of which only three are ``eaten''. Therefore
60 PGBs are left in the low--energy spectrum. Technidilaton
couplings to a generic PGB is given by
$${\cal L}_{\rm GBD} = \left( {1\over 2} \partial_\mu P \partial^\mu P\right)
U^2 - \left({1\over 2} M^2_P P^2\right) {4D\over F_D} \eqno(8)$$
where $P$ denotes a pseudo-Goldstone boson. The first term is the kinetic term
which is rendered scale invariant by multiplying by $U^2$ and the second one
arises from the mass term which is not scale invariant.

In TC, as in any gauge theory, dilatations are broken by the
scale anomaly too.
Other than giving mass to the technidilaton and modifying the technidilaton
self couplings the TC trace anomaly does not
play an important role [3]. In
addition to the TC trace anomaly, there are anomalies arising from the
$ SU(3)_C\times SU(2)_L\times U(1)_Y$ gauge interactions [5]. Since at the
relevant energy scales $(M_W <\!\!<E<\!\!<4\pi F_\pi)$ none of these
interactions are strong we can use the perturbative form of the anomaly
for them. The $ SU(3)_C\times SU(2)_L \times
U(1)_Y $ trace anomaly is given by [5]
$$\Theta^\mu{}_\mu = {\beta(\alpha_s) \over 4\alpha_s} G^{\mu\nu a}
G^a_{\mu\nu} + {\beta(\alpha_W)\over 4\alpha_W} W^{\mu\nu b} W^b_{\mu\nu} +
{\beta(\alpha_Y)\over 4\alpha_Y} B^{\mu\nu} B_{\mu\nu} \eqno(9)$$
where $G^{\mu\nu a}$, $W^{\mu\nu b}$ and $B^{\mu\nu}$ are the $SU(3)_C$,
$SU(2)_L$ and $U(1)_Y$ field strength tensors and $\beta(\alpha_s)$,
$\beta(\alpha_W)$, $\beta(\alpha_Y)$ are the respective $\beta$-functions.
The technidilaton being the Goldstone boson of dilatations couples to the
divergence of its current or to $\partial_\mu D^\mu =\Theta^\mu{}_\mu$
in the form
$${\cal L}_{\rm int} = - {D\over F_D} \Theta^\mu{}_\mu\eqno(10)$$
with $\Theta^\mu{}_\mu$ given by Eq. (9) in addition to the gauge boson
mass terms (in the unitary gauge). Eq. (10) gives new
technidilaton-gauge boson interactions due to the trace anomaly which have
no analog for the Higgs boson of the Standard model.

If the ordinary fermions ({\it i.e.\/} quarks or leptons) obtain their masses
from some mechanism at higher energies like extended technicolor
(ETC) [12], their mass terms contribute to $\Theta^\mu{}_\mu$ by an amount
$\Theta^\mu{}_\mu \hbox{(masses)} = \sum_f m_f \bar{f}f$
where $f$ denotes a fermion (either quark or lepton) and $m_f$ is its mass.
The sum over $f$ runs over all quarks and leptons. From Eq. (10) it follows
that
there are technidilaton-fermion interactions given by
$${\cal L}'_{\rm int} = - {D\over F_D} \Theta^\mu{}_\mu\;\hbox{(masses)} = -
{D\over F_D} \sum_f m_f \bar{f}f \quad . \eqno(11)$$
Eqs. (7) and (11) describe technidilaton interactions with matter
which are very similar to those of the standard Higgs boson.

The properties of the technidilaton can now be compared to those of the
standard Higgs boson $\phi$. In the standard Higgs picture, the
Higgs vacuum expectation value $\langle 0|\phi|0 \rangle = v$ fixes the weak
gauge boson ($W^\pm,
Z^0$) masses and gives the Higgs mass by $m_\phi = \sqrt{2\lambda}\,v$.  Also
Higgs coupling to matter are proportional to ${1/ v}$.  For the
technidilaton the situation is completely different.  According to the picture
described above the gauge boson masses are fixed by $F_\pi$ as
given by Eq. (6). This is similar to the Standard model with $v$
replaced by $F_\pi$ (they have the same value for $r=1$).  On the other hand,
the technidilaton mass, $M_D$, given by Eq. (2) is completely different than
$m_\phi$.  In particular, it is not proportional to either $\sqrt{\lambda}$
(or the coefficient of the $D^4$ coupling) or to $v$ (or $F_\pi$).
$M_D$ is given
by the value of the technigluon condensate given by Eq. (1). Moreover, as we
will see in the following, the cubic and quartic technidilaton self--couplings
can be
strong even when $M_D$ is relatively small ({\it i.e.\/} less than 1~TeV).  In
the Standard model strong scalar interactions imply  a large ($\sim 1$~TeV)
Higgs mass and vice versa [13].

{}From the preceeding arguments one sees that the technidilaton
interactions with fermions and gauge bosons are very similar to those
of the standard Higgs boson. There are a few
important differences however.  First, Higgs couplings are proportional to
${1/ v}$ whereas technidilaton couplings are proportional to ${1/
F_D}$ and not to the analog of ${1/ v}$ which is ${1/ F_\pi}$.  $F_D$
can be estimated by scaling from QCD and using the $1/N$ expansion
which give $F_D \sim 830~\hbox{GeV} \sqrt{{N/ 3}} $ [3].
The same result can also be obtained by finding the energy scale $\Lambda_D$ at
which the technigluon condensate forms and by identifying $\Lambda_D \simeq
4\pi F_D$.
It is seen that for any TC theory with an $SU(N)$ $N>3$ group, $F_D$ is much
larger than $F_\pi$.  Therefore technidilaton couplings to fermions and gauge
bosons are much weaker (by a factor of $v/F_D \simeq \sqrt {0.3/N}$) than
those of the standard Higgs boson even though they
both have the same form.

Moreover, because of the scale anomaly, when quantum effects are
taken into account one gets additional technidilaton-gauge boson interactions
given by Eq. (10). Consider the interactions due to the anomaly for the
$SU(2)_L$ gauge bosons.  From Eqs. (9) and (10)
$$\eqalignno{{\cal L}_{\rm int} (DW^+W^-) &= - {D\over F_D} \
{\beta(\alpha_W)\over 4\alpha_W} W^{\mu\nu b} W^b_{\mu\nu}  \cr
&\simeq  {D\over F_D} W^{+\mu} W^-_\mu {\beta(\alpha_W)\over 2\alpha_W}
p_1\cdot p_2 &(12) \cr}$$
where $p_1$ and $p_2$ are the four-momenta of the two gauge bosons.  It is
seen that this term is very small at low energies ({\it i.e.\/} $E<\!\!<5M_W$)
because of the factors $p_1\cdot p_2$ and ${\beta(\alpha_W)\over \alpha_W}$.
Therefore the technidilaton-weak gauge boson interactions
are not modified by much for
$E<\!\!<M_W$.  For higher energies $5M_W < E <\!\!< 4\pi F_\pi$, Eq. (12)
becomes appreciable compared to the classical interactions which are
proportional to the gauge boson masses.  For gauge bosons with three-momenta
$p \sim 0.5$~TeV, the anomaly term is 30\% of the classical interaction for
transverse gauge bosons given by
Eq. (7).  Note that Eq. (12) modifies the couplings to
transverse gauge bosons only. Since $\beta(\alpha_{TC})$ is negative
the anomaly decreases the technidilaton-transverse gauge boson couplings.
The technidilaton does not couple to gluons or photons classically (as the
standard Higgs boson) since they are massless. But the $SU(3)_C$ and
$U(1)_{EM}$
trace anomalies induce technidilaton couplings to photons and gluons similar
to Eq. (12). Even though these interactions are weak, they get stronger as the
c.o.m. energy increases. The anomalous coupling of $D$ to gluons
plays a very important role in $D$ production due to gluon fusion. We will see
that as a result of this effective coupling the coupling of $D$ to
gluons is not suppressed relative to the Higgs coupling.
This makes $D$ production by gluon fusion the only realistic production
mechanism. The anomalous coupling of $D$ to photons will play a mojor role in
observing the $D$'s produced at the SSC in their rare decay modes $D \to 2
\gamma$ for the mass range $40~GeV<M_D<2M_P$.

At energies much higher than gauge boson masses ($E>\!\!>M_W$) the
longitudinal gauge bosons behave as technipions.  Therefore the technidilaton
interactions with them are not given by Eq. (7) but by
$${\cal L}_{\rm DP} = {F^2_\pi\over 4} \Tr \left[ \left(
\partial_\mu\Sigma\right) \left(\partial^\mu \Sigma\right)^\dagger
 \right] U^2 \ \ .\eqno(13)$$
In terms of the technipions $\Pi^a$ to lowest order
$${\cal L}_{\rm DP} = \left( \partial_\mu \Pi^+ \partial^\mu \Pi^- + {1\over
2}\partial_\mu \Pi^0 \partial^\mu \Pi^0\right) U^2 \ \ .\eqno(14)$$
This is very similar to what happens in the Standard model.  For $E>\!\!>M_W$,
the longitudinal polarization vector for the gauge bosons
$\epsilon^L_{\mu}= {1\over M_W} (k,0,0,E)
\simeq {k_\mu / M_W} + {\cal O} \left({M_W/ E}\right) $
for a gauge boson with energy $E$ and momentum $k_\mu = (E,0,0,k)$.  Therefore
$$\left( M^2_W W^+_L W^-_L + {1\over 2} M^2_Z Z^{02}_L\right) {2\phi\over v}
\simeq \left[ \partial_\mu \phi^+\partial^\mu \phi^- + {1\over 2}
\partial_\mu (Im \phi^0) \partial^\mu (Im \phi^0) \right] {2\phi\over v}
\eqno(15)$$
where $\phi^+$, $\phi^-$ and $Im \phi^0$ are the scalars which become
$W^+_L$, $W^-_L$ and $Z^0_L$.

The cubic and quartic technidilaton interactions can be obtained from Eq. (5)
and the TC anomaly term [3] applied to Eq. (10). We get
$${\cal L} (D^3, D^4) \simeq \left( {D^3\over F_D}
+ {D^4\over F^2_D}\right)  \left( {s\over 2} + {M^2_D\over 3}\right) \quad .
\eqno(16)$$
We see that $D^3$ and $D^4$ couplings are momentum-dependent unlike $\phi^3$
and $\phi^4$ couplings given by the Standard model. For $s<\!\!< M^2_D$,
$D^3$, $D^4$
couplings are of the order of $\phi^3$, $\phi^4$ couplings whereas
for $s>\!\!> M^2_D$ they are
much larger. (The energy range $s>\!\!>M^2_D$ might be
beyond the range of validity of the effective TC Lagrangian if $M_D >1$~TeV.)
In the standard model if $m_\phi <\!\!<1$~TeV, the standard Higgs will be
weakly interacting and vice versa. For $D$, even when $M_D<\!\!<1$~TeV, the
cubic and quartic self--couplings grow with $s$ and become strong around
$s\sim 2$~TeV$^2$ (at which the effective Lagrangian may or may not be valid).

There are also interactions with higher orders of $D$ ({\it e.g.\/} $D^5, W^+
W^- D^3$, {\it etc.\/}) which arise from the expansion of $U$ in terms of $D$.
 These are suppressed by at least a factor of ${s/ (4\pi F_D)^2}<\!\!< 1$
relative to the terms discussed above and therefore will be neglected.

It is seen that the major difference between the Higgs $\phi$ and the
technidilaton $D$ is that, interactions of $D$ are suppressed by a factor of
${v/ F_D}$ relative to those of $\phi$.  Therefore if a scalar boson is
observed
it will be relatively easy to distinguish between these two alternatives
by observing
their production and decay rates. Neglecting the effects of the anomaly, one
would naively expect both the production and decay rates for the technidilaton
to be suppressed by a factor of $v^2/F_D^2 \simeq 0.3/N$ relative to those of
the standard Higgs boson.
\bigskip
\centerline {\bf 3. Technidilaton decays and production mechanisms}
Since the technidilaton's interactions are very similar to those of the Higgs
boson, its decay channels are the same as those for the Higgs which are
mainly into fermion--antifermion pairs, $W^+
W^-$ or $Z^0Z^0$. (In addition, $D$ can decay into light PGBs if this is
kinematically allowed. If the number of decay channels into PGBs is large, the
branching ratios for the other channels might be reduced significantly.) The
major difference between the two cases is the suppression
factor $v/F_D$ in the technidilaton couplings. Therefore $D$ decay rates
are suppressed by $v^2/F_D^2 \sim 0.3/N$ with respect to those of the Higgs
boson. As for the Higgs boson [14], for $M_D<2M_W$ the dominant decay is into
$\bar b b$ since $b$ is the heaviest particle available. The decay rate into
a fermion--antifermion pair is
$$\Gamma(D \to \bar f f)={{N_cM_Dm_f^2}\over{8\pi F_D^2}}\left(1-{4m_f^2\over
M_D^2} \right)^{3/2}\quad . \eqno(17)$$
For $M_D>2M_Z$ the dominant decay becomes that into vector bosons $W^+W^-$
and $Z^0Z^0$ with the decay rates
$$\Gamma(D \to W^+W^-)={M_D^3\over{16\pi F_D^2}}(1-x_W^2)^{1/2}\left(1-x_W+
{3\over4}x_W^2 \right) \eqno(18)$$
where $x_W=(4M_W^2/M_D^2)$ and
$$\Gamma(D \to Z^0Z^0)={M_D^3 \over{32\pi F_D^2}}(1-x_Z^2)\left(1-x_Z+{3\over4}
x_Z^2 \right) \eqno(19)$$
where $x_Z=4M_Z^2/M_D^2$.

 If $M_D>2m_t$, $D$ decays into the top quark but the main decay
channels are always those into the vector bosons at this mass range (even
though $m_t>M_W,M_Z$). This is easily seen by comparing Eq. (17) and
Eqs. (18,19).
We saw in Section 2 that the $D$ couplings to the transverse gauge
bosons are modified due to the gauge anomalies (Eq.(12)). One might hope that
these modifications would be appreciable for a range of technidilaton masses
and could be observed in the ratio of transverse to longitudinal gauge bosons
in $D$ decays. Unfortunately, these corrections are negligible for $M_D<5M_Z$,
at
which 99$\%$ of the decay products are already longitudinal gauge bosons [15].
Therefore, in practice, the effects of the extra couplings due to the gauge
anomalies will not be important in $D$ decays.

In addition to these decay channels there are two rare decays of $D$ which are
important. These are $D \to 2\gamma$ and $D \to 2g$ which are analogous to
the rare Higgs decays in these channels [14]. For the technidilaton decays
these
proceed not only through the one--loop diagrams as for the Higgs but also (and
mainly)
through the technidilaton couplings due to the QED and QCD trace anomalies
given by Eq. (9). In addition to the ordinary fermions, the technifermions
also contribute to the $U(1)_Q$, $SU(3)_C$ and $SU(2)_L$ $\beta$--functions
since in general they carry electric, color and weak hypercharges [4].
As a result, processes which depend on the $\beta$--functions such as $D$
decays into $2g$ or $2\gamma$ and production by gluon fusion will be model
dependent.
For concreteness we consider two
models: Model 1 with $n_T$ technifermion generations and TC gauge group
$SU(N)_{TC}$. The technifermion quantum numbers under
$SU(3)_C \times SU(2)_L \times U(1)_Y$ are identical to those of ordinary
fermions in order to cancel gauge anomalies.
In Model 2 there are $r$ $SU(2)_L$ technifermion doublets which are color
singlets with the TC gauge group $SU(N)_{TC}$. The $U(1)_Y$ charges are 0,${1
\over 2}$ and $-{1\over 2}$ for $T_L$, $A^i_R$ and $B^i_R$ respectively so as
to cancel all anomalies.

The decay $D \to 2\gamma$ takes place mainly through the anomaly coupling
$${\cal L}_{an}=-{D\over F_D}{\beta(\alpha_{EM})\over{4\alpha_{EM}}}F^{\mu\nu}
F_{\mu\nu} \eqno(20)$$
where $F^{\mu\nu}=\partial^\mu A^\nu-\partial^\nu A^\mu$ is the
electromagnetic field strength tensor.
For Models 1 and 2 the electromagnetic $\beta$--functions are given by
$$\beta(\alpha_{EM})={1\over4\pi}\left({32\over9}(n_g+n_TN)\right)
\alpha_{EM}^2 \eqno(21a) $$
and
$$\beta(\alpha_{EM})={1\over4\pi}\left({32\over9}n_g+{Nr \over 3}\right)
\alpha_{EM}^2 \eqno(21b) $$
respectively where the terms with $n_TN$ and $Nr$ are the contributions of the
technifermions and $n_g$ is the number of fermion generations which is three.
The $\beta$--functions of $SU(2)_L$ for the two models are given by
$$\beta(\alpha_W)={1\over8\pi}\left({4\over3}(n_g+n_TN)-{44\over6}\right)
\alpha_W^2 \eqno(22a)$$
and
$$\beta(\alpha_W)={1\over8\pi}\left({4\over3}n_g+{Nr\over 3}-{44\over6}\right)
\alpha_W^2 \quad . \eqno(22b)$$
Note that for three generations, $\beta(\alpha_W)$ changes sign
for $n_TN>3$ or $Nr>10$. In the following we consider the two cases with
$n_T=3$ in Model 1 and $r=6$ in Model 2 which correspond to a $D$ with $M_D=
100~GeV$ and $M_D=200~GeV$ respectively. All the results can easily be
generalized to cases with $n_T$ technifermion generations or $r$ doublets
by taking into account the $n_T$ or $r$ dependence of $\beta$--functions,
$M_D$ and $F_D$. From Eqs. (21a,b) we get by substituting $n_g=3$, $N=6$,
$n_T=3$ for Model 1 and $r=6$ for Model 2 the decay rate [14]
$$\eqalignno{\Gamma(D \to2\gamma) &\sim c_1{{\alpha_{EM}^2M_D^3}\over
{\pi^3F_D^2}}&(23) \cr
&\sim 10c_1{v^2\over{F_D^2}}\Gamma(\phi \to 2\gamma) &(24) \cr}$$
where $c_1 \sim3$ and $c_1 \sim0.3$ in the first and second cases
respectively. We see from Eqs. (21a,b) that the dependence of
$\beta(\alpha_{EM})$ on $n_TN$ or $r$ is very strong.
The relation between the branching ratios for $D$ and the Higgs boson decays
into $2\gamma$ is
$$Br(D \to 2\gamma) \sim 10c_1 Br(\phi \to 2\gamma) \quad . \eqno(25)$$
We see that $Br(D \to 2\gamma)$ is enhanced by factors of $\sim 30$ and $\sim
3$
for the two cases considered respectively.

The other rare decay, which is even more important, is $D \to 2g$ which takes
place due to the technidilaton coupling to the QCD trace anomaly
$${\cal L}_{an}=-{D\over F_D}{\beta(\alpha_s)\over{4\alpha_s}}G^{\mu\nu a}
G^a_{\mu\nu}\quad  a=1,\ldots,8 \eqno(26)$$
where $G^{\mu\nu a}$ is the QCD field strength tensor and the QCD
$\beta$--function to lowest order in Model 1 is given by
$$\beta(\alpha_s)=-{1\over2\pi}\left(11-{4\over3}(n_g+n_TN)\right)\alpha_s^2
\eqno(27a)$$
whereas in Model 2
$$\beta(\alpha_s)=-{1\over2\pi}\left(11-{4\over3}n_g\right)\alpha_s^2
\eqno(27b)$$
since technifermions of Model 2 do not carry color.
Note that $\beta(\alpha_s)$ changes sign for $n_TN>5$.
For the two cases considered above, with $n_g=3$, $N=6$, $n_T=3$ and $r=6$
(assuming $M_D=m_{\phi}$), including the
contribution of the quark loop diagrams, one gets [14]
$$\eqalignno{\Gamma(D \to2g) &\sim c_2{{\alpha_s^2M_D^3}\over{\pi^3F_D^2}}&(28)
\cr &\sim 60c_2{v^2\over{F_D^2}}\Gamma(\phi \to 2g) &(29) \cr}$$
where $c_2 \sim2$ and $c_2 \sim 1/3$ for the two models respectively.
We see that this decay rate, too, is enhanced in the first case by a factor of
$\sim 35/N$ compared to the respective Higgs decay. In the second case, the
decay rate
is neither enhanced nor suppressed relative to that of the Higgs. The
importance of this result is apparent: the effective technidilaton coupling
to two gluons is not suppressed (or even enhanced) whereas all the other
couplings playing a major role in
technidilaton production are suppressed by a factor of $\sqrt{0.3/N}$. As
a result, gluon fusion will be the only realistic production mechanism for $D$.

If there are more than one technifermion doublet, there will be a number of
PGBs in the spectrum. If $M_D>2M_P$ where $M_P$ is the
PGB mass then $D$ can decay into those particles as well.
{}From the generic coupling of the technidilaton to the PGBs
one obtains the decay rate
$$\Gamma(D \to PP)={M_D^3\over{16\pi F_D^2}}(1-x^2)^{1/2}(1-6x+9x^2)
\eqno(30)$$
where $x=4M_P^2/M_D^2$. These decays are as important as those into gauge
bosons and in general they will compete. The decays into PGBs are
expected to be important only for a technidilaton with $M_D>100~GeV$
since the lightest PGBs are expected to have masses of
$\sim 50~GeV$ [4].

Since the technidilaton couplings are suppressed by a factor of $\sqrt{0.3/N}$
relative to those of the Higgs boson, the total width of $D$ will be much
smaller unless $D$ mass is such that decays into PGBs are allowed. For a heavy
technidilaton (i.e. with $M_D>1~TeV$ and $x_W$, $x_Z$, $x << 1$) one gets
$$\Gamma_{tot} \sim 60~GeV \times \left(M_D\over{1~TeV}\right)^3 \left(3\over
N \right) \quad . \eqno(31) $$
It is seen that for $N=6$ $D$ is about fifteen times narrower than a Higgs
boson of the same mass [14]. Therefore, even for $M_D \sim 2~TeV$, $D$ will
be a clearly
defined peak in the cross section contrary to the Higgs boson which for
$m_{\phi}>1~TeV$ is a wide resonance. For the mass range $2M_Z<M_D<1~TeV$,
there
are decays into PGBs which may increase $\Gamma_{tot}$ by a factor of up to 40.
This enhancement depends on the specific TC model considered which gives the
number of light PGBs and their masses. For example, in Model 2 with $r=6$ and
$N=6$ (which give $M_D \sim 200~GeV$) one gets $\Gamma_{tot} \sim 10~GeV$.
In the mass range $2M_A<M_D<2M_Z$, ($A$, a generic axion of TC models, is the
lightest PGB. We will discuss PGBs in TC models in more detail later.) $D$
width is much larger than that
of the Higgs since $D$ can decay into the light axions with a decay rate that
is much larger then that into $\bar b b$.
For a light $D$ with $M_D<2M_A \sim 100~GeV$, $D$ width is smaller then
that of the Higgs by the factor $v^2/F_D^2$ since a light $D$ cannot decay into
PGBs.

The main production mechanisms for both the Higgs boson and the technidilaton
are associated production (with a $Z^0$) [16], vector boson fusion [17] and
gluon fusion [18].
We will see that for $D$ the only production mechanism  which is not
suppressed is gluon fusion.
This fact makes gluon fusion the only realistic technidilaton production
mechanism. We consider each mechanism seperately for $D$.

a) Associated Production: This takes place in an $e^+e^-$ collider by
$e^+e^- \to Z^0 \to Z^0 D$ or in a $ p p$ collider by $\bar q q \to Z^0 \to
Z^0 D$. Since the technidilaton coupling to $Z^0$ is
the same as those of the Higgs except for a suppression factor of $v/F_D$
we immedeately obtain for $M_D=m_{\phi}$
$$\sigma(e^+e^- \to Z^0 \to Z^0 D)=(v^2/F_D^2)\sigma(e^+e^- \to Z^0 \to Z^0
\phi) \eqno(32) $$
where the cross section for the Higgs boson is given by [16]
$$\sigma(e^+e^- \to Z^0 \to Z^0 \phi)={{\pi \alpha_{EM}^2
\lambda^{1/2}[\lambda+
12sM_Z^2][1+(1-4sin^2{\theta_w})^2]}\over{192s^2 sin^4{\theta_w}cos^4{\theta_w}
(s-M_Z^2)}} \eqno(33)$$
with $\lambda=(s-m_{\phi}^2-M_Z^2)^2-4m^2_{\phi}M_Z^2$ and $k=\sqrt\lambda/2
\sqrt2$ is the center of mass momentum of either $Z^0$ or $\phi$.
The cross section $\sigma$ as a function of c.o.m. energy $E$ is given by
Fig. 1 for $M_D=30,60,100~GeV$. Note that the peak decreases rapidly with
increasing $M_D$.
At very high energies (i.e. $s>>m_{\phi}^2,M_Z^2$) this cross section is
proportional to $1/s$ and therefore decreases with energy. This channel is
the dominant production mechanism in $e^+e^-$ colliders at low energies.

The cross section for the $\bar q q $ case can be easily obtained by
substituting the appropriate values for $g_V$ and $g_A$ for the quarks in the
scattering amplitude.

b) Vector Boson Fusion: This, too, can take place in both $e^+e^-$ and $ p
p$ colliders. Once again the suppression of $DVV$ couplings, where
$V$ is a generic vector boson, by $v/F_D$ relative to $\phi VV$ couplings gives
for $M_D=m_{\phi}$
$$\sigma(e^+e^- \to D l \bar l)=(v^2/F_D^2)\sigma(e^+e^- \to \phi l \bar l)
\eqno(34)$$
where the cross section for the Higgs boson is given by [17]
$$\sigma(e^+e^- \to \phi l \bar l)\simeq{{g_{VV\phi}^2}\over{16\sqrt6
\pi^3 M_V^4}}(c_1+c_2) ln{\left({s\over{m_\phi^2}\right)}} \eqno(35)$$
with $c_1=g_L^2g_L^{\prime2}+g_R^2g_R^{\prime2},\quad c_2=g_L^2g_R^{\prime2}+
g_R^2g_L^{\prime2}$. Here $l$ and $V$ denote either $e^-$ and $Z^0$ or $\nu_e$
and $W^+$. $M_V$ is either $M_Z$ or $M_W$ and $g_{VV\phi}=(M_V^2/v)$.
We neglected the enhancement of $D$ couplings to longitudinal gauge bosons
which will increase $\sigma(e^+e^- \to \phi l \bar l)$ slightly.
This cross section as a funtion of c.o.m. energy $E$ is given in Fig. 2. It
is seen that $\sigma$ decraeses with increasing $M_D$
but increases with energy albeit
logarithmically. At very high energies to be determined later, vector boson
fusion overtakes associated
production and becomes the dominant production mechanism in $e^+e^-$ colliders.
The cross section for the $\bar q q$ collision can be obtained from Eq. (35)
by substituting the relevant $g_{L,R}$ for quarks.

c) Gluon Fusion: This takes place only in $ p p$ collisions. It was found
in the last section that the effective technidilaton coupling to two gluons
is not suppressed relative to the Higgs coupling. Then the cross section [18]
$$\sigma( p p \to D+X)={{\pi^2 \Gamma(D \to 2g)}\over{8M_D^3}}\tau{d{\cal L}
\over d\tau} \eqno(36)$$
with
$${d{\cal L}\over{d\tau}}=\int^1_0dx_1 \int^1_0dx_2 \delta(\tau-x_1x_2)f_p(x_1)
f_{ p}(x_2) \eqno(37)$$
is not suppressed either.
Here $\tau=m_D^2/s$, $x_1$ and $x_2$ are the gluon momentum fractions in $p_1$
and $ p_2$, $f_p(x_1)$ and $f_{\bar p}(x_2)$ are the gluonic momentum
distribution functions in $p_1$ and $ p_2$. These distribution functions are
obtained from the measured low--energy values by the use of Altarelli--Parisi
equations [19]. Since $\Gamma(D \to 2g)$ is model dependent as we saw before,
from Eq. (36) we see that $D$ production cross section is model dependent too.
In terms of the cross section for the Higgs boson one gets
for a $D$ and $\phi$ of the same mass
$$\sigma(p  p \to D+X)\simeq {c_3\over N}\sigma(p p \to \phi+X)
\eqno(38)$$
where $c_3 \sim 35$ and $c_3 \sim 6$ in cases 1 and 2 respectively.
$\sigma(p p \to D+X)$ for some values of collider energy and for
$M_D=100~GeV$ is given by Table 1 for the two models. It is seen that the
cross section $\sigma$ increases in both cases as the c.o.m. energy
increases as expected. $\sigma_1$ is about six times larger than $\sigma_2$
since $\beta(\alpha_s)$ is larger in Model 1 (for $M_D=100~GeV$).

\bigskip
\centerline{\bf 4. Prospects for technidilaton detection}
In this section we discuss the prospects for producing and observing the
technidilaton at present and future accelerators. The production and
observation of $D$ at $e^+e^-$ colliders is extremely difficult compared to
the Higgs boson. As we saw in the last section, the cross
section for $D$ production is smaller than that for $\phi$ by a factor of
$\sim 0.3/N$ for a $SU(N)_{TC}$ gauge group. Since the background events are
not suppressed, for $D$ the signal to background ratio will be very small.

A point worth clarifying is the presence of light PGBs.
If they exist, the decay channels $D \to PP$ will be important in addition to
those into vector bosons.
As will be seen below, $e^+e^-$ colliders can only find relatively light
(i.e. $M_D<40~GeV$) technidilatons. From the formula $M_D \sim (7.5/Nr)~TeV$
we see that in order to get a light $D$, $r$ (or $4n_T$) must be $O(10-15)$.
In this case one expects to have a large number of PGBs. First, consider
Model 1 with three technifermion generations.
In this case most of the PGBs are color octets or triplets
with masses $M_O \sim 260 \sqrt{4/N}~GeV$ and $M_T \sim 170 \sqrt{4/N}~GeV$
respectively [4]. In addition, in every generation, there are four light axions
with $M_A \sim 50~GeV$. These get most of their masses from ETC interactions
[4]. In Model 2, on needs $r \sim 10-15$ in order to get a light $D$. Then
there
will be a very large number (a few hundred) of light axions with $M_A \sim
50~GeV$ as before. The exact number of PGBs and their masses are highly model
dependent.
For our purposes, the generic results stated above will be enough.

We see that a light $D$ (with $M_D<100~GeV$) cannot decay into PGBs even though
there may be a large number of them. Therefore, the existence of PGBs will not
affect the results obtained for $e^+e^-$ colliders at which only light $D$s can
be observed as we will see below. For
$100~GeV<M_D<200~GeV$, $D$ can decay into the axions whose number we take to be
$\sim 10$ in Model 1 and $\sim 130$ in Model 2 (since for this mass range we
need either $n_T=2$ or $r=6$). For $M_D>300~GeV$, in Model 1,
$D$ will decay into the color triplets (and possibly octets) too. As a result,
the number of decay channels will increase
significantly which will reduce the branching ratios  for the other
conventional decays. In Model 2 on the other hand, all PGBs are light so that
there are no thresholds above $\sim 100~GeV$.The effects of the PGBs will be
particularly important
in the $D$ mass range to be relevant at the SSC as we will see later.

We consider the different $e^+e^-$ colliders and the SSC seperately.

a) LEP 1/SLC: These are $e^+e^-$ colliders with $\sqrt s \sim 100~GeV$. The
principal technidilaton production at LEP 1 energies is associated production
as can be seen from Fig. 3. In this process
the intermediate $Z^0$ is real but the final one is not. The final $Z^0$
can be tagged by its decay into either $e^+e^-$ or $\nu \bar \nu$.
$D$ decays mainly into $\bar b b$ since it is light i.e. $M_D<2M_W$. With a
luminosity of $L=10^{31} cm^{-2}sec^{-1}$, in one year ($10^7$ sec) one expects
to have $10^6$ $Z^0$'s produced at LEP 1 [20]. For the cross section $e^+e^-
\to
D Z^0 \to De^+e^-$, using the results $Br(Z^0 \to e^+e^-) \sim 3\%$ and
$$Br(Z^0 \to D \bar f f )<(3/N)\times 10^{-3}Br(Z^0 \to \bar f f) \eqno(39)$$
one gets
$$Br(e^+e^- \to D Z^0 \to D e^+e^-) \sim 10^{-5} \quad . \eqno(40)$$
This corresponds to 10 events per year for a massless $D$ and the number of
events decreases rapidly with increasing $M_D$ [14]. For $M_D=20~GeV$ one gets
only about one event per year
which clearly inadequate. So this mode can only find or rule out  very light
$D$'s with a mass up to $\sim 10~GeV$. In the $\nu$--channel with $Z^0 \to
\bar \nu
\nu$ with a branching ratio $Br(Z^0 \to \bar \nu \nu) \sim 19 \%$ there are
many more events. Unfortunately this channel has a large background due to the
usual $e^+e^- \to \gamma \to \bar b b$ events [20]. As a result, we conclude
that
LEP 1 can detect or rule out a technidilaton with $M_D\sim 10~GeV$ compared to
the lower bound on the Higgs boson, $m_{\phi} > 60~GeV$ [21].

b) LEP 2: This will be an upgraded version of LEP 1 with $\sqrt s \sim
200~GeV$.
Here too the main production mechanism is associated production for
technidilaton masses up to $\sim 100~GeV$ as can be seen from Fig. 4. For
larger $M_D$, $\sigma$ for both production mechanisms are negligible so the
effects of vector boson fusion are not important.
 Now, the
intermediate or the final $Z^0$ is virtual if $M_Z<\sqrt s < M_Z+\sqrt2 M_D$
or $\sqrt s >M_Z+\sqrt2 M_D$. The cross section for this event is given by
Eq. (35) and Fig. 4. One gets 48 fb for $M_D=20~GeV$ and 16 fb for
$M_D=100~GeV$. For a
luminosity $L=5\times 10^{31} cm^{-2} sec^{-1}$, in one year this corresponds
to 24 and 10 events respectively. Using the results of the ALEPH collaboration
[22]
for the Higgs, one finds that for the $Z^0D \to \bar \nu \nu \bar b b$ channel
there are five events with two in the background after all the cuts and
selection rules are imposed for $M_D=40~GeV$ [20]. For larger masses the
signal to background ratio is too small.

To escape the large background one can look for the leptonic channels $Z^0D
\to e^+e^- \bar b b$ or $Z^0D \to \mu^+ \mu^- \bar b b$. Although these have
small bachgrounds, the rates are also smaller since $Br(Z^0 \to l^+l^-) \sim
3\%$. For example, from ALEPH collaboration [22] results one finds $\sim 1.8$
events with 0.2 backgrounds for $M_D=40~GeV$. The $Z^0 \to \bar q q$ mode
is not useful because of the large QCD background from 4--jet events.
We conclude that LEP 2 can find or rule out technidilatons with a mass up to
$\sim 40~GeV$. This can be compared to the expected bounds on the Higgs boson
mass, $m_{\phi} \sim 85,92,110~GeV$ for LEP 2 center of mass energies $175,190,
210~GeV$ respectively [23].

c) TLC: This is a planned $e^+e^-$ collider with $\sqrt s=1~TeV$ and $L=10^{33}
cm^{-2}sec^{-1}$. At these energies the dominant $D$ production mechanism is
vector boson fusion as can be seen from Fig. 5. From the cross sections
for associated production and
vector boson fusion given by Eq. (33) and Eq. (35) we find that the latter
overtakes the former around $\sqrt s=0.1~TeV+0.9M_D$. This means that for
$\sqrt s=1~TeV$, vector boson fusion is dominant for up to $M_D\sim 1~TeV$.
The cross section Eq. (35) gives
$$\sigma(e^+e^- \to D \bar \nu_e \nu_e) \simeq 6.5fb\times ln(s/M_D^2) \quad .
\eqno(41)$$
For $\sqrt s=1~TeV$, $M_D=100~GeV$, a one year run gives about 250 events.
These will be completely washed out by the huge amount of background events
such as $e^+e^- \to \bar b b$ (1000 events), $e^+e^- \to W^+W^-$ (27000
events), $e^+e^- \to Z^0Z^0$ (15000 events) etc. [20] Even after all the cuts
and
selection criteria the signal to background ratio is expected to be very small
[20].

It is seen that high--energy $e^+e^-$ colliders are not very useful for
finding the technidilaton. Of course they are very important for Higgs
production. TLC can produce a standard Higgs boson with a mass up to 150 GeV
[20].
The difference between the technidilaton and the Higgs boson is the factor
$0.3/N$ in the production cross section (for $e^+e^-$ colliders) by which the
technidilaton production is suppressed. The logarithmic dependence on energy
in Eq. (41) makes it impossible to compensate this factor by going to higher
energies. Conversely, if at a given c.o.m. energy $\sqrt s$, no technidilaton
is
observed the upper bound on $M_D$ will be much smaller than that for $m_{\phi}$
because of the logarithmic dependence on the masses in Eq. (41). These results
are model independent since the relevant production and decay mechanisms in
$e^+e^-$ colliders are not model dependent. In addition, PGBs are also
irrelevant in these cases.

d) SSC: This is a $ p p$ collider with $\sqrt s=40~TeV$ and luminosity
$L=10^{33}cm^{-2}sec^{-1}$. The main problem with a $ p p$ collider is the
large QCD backgrounds and disentangling the relevant subprocess from the
$p p$ collision. A priori all the $D$ production mechanisms discussed in the
previous section take place at a $ p p$ collision. At $\sqrt s=40~TeV$,
the parton subprocess has a center of mass energy of a few TeV. At these
energies associated production is completely negligible.

Since $D$ production by vector boson fusion is suppressed by $0.3/N$ but
production by gluon fusion is not, the dominant $D$ production mechanism at
the SSC will be gluon fusion for all $m_t$. For the Higgs boson, these two
processes may be
comparable at these energies depending on $m_t$ [15]. The cross section for
$D$ production by gluon fusion is given by Eq. (36)
$$\sigma( p p \to D+X)={{\pi^2 \Gamma(D \to 2g)}\over{8M_D^3}}\tau{d{\cal L}
\over d\tau} \eqno(36)$$
with $\Gamma(D \to 2g)$ given by Eq. (29). The gluonic momentum distributions
must be evaluated at $Q^2=M_D^2$. This is done by taking the measured function
$f_p(x,Q^2=5~GeV^2)$ [24] and using the Altarelli--
Parisi equation for the evolution of $f_p$.

The cross section $ \sigma(p p \to D+X)$ at the SSC for some values of
technidilaton mass $M_D$ in the two models is given in Table 2. In Model 1,
the technifermions have color and their contribution to $\beta(\alpha_s)$
changes with $M_D$ since the value of $M_D$ fixes $n_T$ which enters
$\beta(\alpha_s)$. As $M_D$ increases, $n_T$ and therefore $\beta(\alpha_s)$
decrease. We see that $\sigma_1$ decreases very rapidly as $M_D$ increases
from 100 GeV to 300 GeV because of the decreasing phase space and $\beta(
\alpha_s)$ which gives the $Dgg$ coupling. Note that
$M_D>500~GeV$ cannot be accomodated in Model 1. In Model 2, the technifermions
do not carry color and as a result $\beta(\alpha_s)$ is constant. We see that
as $M_D$ increases $\sigma$ decreases as expected from phase space
considerations.

Consider the mass range $2M_Z<M_D$ first. For $M_D=200~GeV$,
Eq. (36) gives cross sections of 40 pb and 27 pb in the two models
respectively. In a year, this gives $4 \times 10^5$ and $2.7 \times 10^5$
$D$s at the SSC. The clearest signal of a technidilaton without a large
QCD background
is $D \to Z^0Z^0 \to l^+l^-l^+l^-$ with $l$ denoting either $e^-$ or $\mu^-$.
Note that for $D$ with $2M_Z<M_D<300~GeV$ one needs $n_T=2$ in Model 1
in which there are $\sim 10$ axions with $M \sim 50~GeV$ and
$\sim 50$ color triplets with $M \sim 150~GeV$ [4]. We will take these channels
into account when the respective thresholds are passed.
With $Br(Z^0 \to l^+l^-) \sim 6\%,Br(D \to Z^0Z^0) \sim 1/13$ a sum over all
combinations gives
$$Br(D \to Z^0Z^0 \to l^+l^-l^+l^-) \sim 2.7 \times 10^{-4} \eqno(42)$$
Then about 110 $D$s will be observed in this mode in one year.
$Br(D \to Z^0Z^0) \sim 1/13$ and not $\sim 1/3$ as usual because of the $D$
decay channels into the $\sim 10$ light axions.
In Model 2 with $M_D>2M_Z$, $r=6$ and the number of light axions is $\sim 130$.
As a result, $Br(D \to Z^0Z^0) \sim O(10^{-2})$ and the number of $D$s observed
is reduced to about 8.
Since $\Gamma(D \to 2g) \sim M_D^3$, the only dependence of the production
cross section Eq. (42) comes from the factor $\tau {d{\cal L}\over{d\tau}}$.
Even
though $\tau$ is quadratic in $M_D$, $\tau {d{\cal L}\over{d\tau}}$ decreases
when $M_D$ increases because $f_p(x)$ drops sharply as $x$ increases [22].

For $M_D \sim 300~GeV$, in Model 1, one gets a cross section of $\sim 0.08~pb$
which corresponds to less then one $D$ observed in the mode given by Eq. (42).
For $M_D>300~GeV$, the branching ratio in Eq.(42) decreases further by a
factor of 4--5
due to the additional decay channels into color triplet PGBs. Since $\sigma_1$
decreases so rapidly in the range $200~GeV<M_D<300~GeV$, we get $M_D<230~GeV$
as a loose upper
bound for the mass of an observable $D$. So in Model 1
one can search the mass range $2M_Z<M_D<230~GeV$. In Model 2, there are not too
many $D$ events for $M_D>200~GeV$ due to the already large number of decay
channels so only the narrow range $2M_Z<M_D<230~GeV$ can be searched.

For the small mass range $2M_W<M_D<2M_Z$, the decay channel $D \to W^+W^- \to
l^+ \bar \nu l^- \nu$ is not useful because of the large backgrounds.
Fortunately, one
can look at the decay channel $D \to Z^0Z^{0*} \to l^+l^-l^+l^-$ where $Z^{0*}$
is a virtual $Z^0$. This mass range can be easily searched in Model 1 whereas
in Model 2, the situation is marginal due to the fewer number of events.

If $M_D<2M_W$, then the dominant decay would be either into $\bar b b $ for
$M_D<2M_A \sim 100~GeV$ or into $\bar A A$ for $2M_A<M_D<2M_W$.
Unfortunately the $D \to \bar b b$
mode has a large QCD background which renders the SSC useless in this
mode and mass range. However, one can look at the rare decay mode $D \to
2\gamma$ which does not have such a large QCD background. For $M_D<2M_A$
the branching ratio is
$$Br(D \to 2\gamma) \sim 0.27c_1\alpha^2 \left(M_D\over m_f \right)^2
\eqno(43)$$
which is larger than Eq. (42). The numerical factor $c_1$ is fixed by $\beta
(\alpha_{EM})$ for a given mass $M_D$ in a given model. In general as $M_D$
increases $c_1$ decreases
since the larger $M_D$ the smaller $n_T$ or $r$. In Model 1,
$Br(D \to 2 \gamma)$ increases from $4.3 \times 10^{-3}$ for $M_D \sim 40~GeV$
to
$1.7 \times 10^{-2}$ for $M_D \sim 100~GeV$.
Assuming that the gluon luminosity increases with decreasing $M_D$ or $x$,
these give $5 \times 10^4$ and $1.4 \times 10^5$ $D$s  for
$M_D=40,100~GeV$ respectively.
In Model 2, on the other hand, $Br(D \to 2 \gamma)$ increases from $1.5 \times
10^{-3}$ to $3.2 \times 10^{-3}$ as $M_D$ increases from $40~GeV$ to $100~GeV$.
These give 3800 and 5100 $D$ events respectively.
These should be
compared to the QCD background (from $\bar q q \to 2\gamma$ and $g g \to
2 \gamma$) which give 30000 and 10000 events respectively. Therefore we
conclude that the mass range $40~GeV<M_D<100~GeV$ can also be searched at the
SSC by the rare decay $D \to 2 \gamma$. A standard Higgs boson with mass less
then $80~GeV$ cannot be detected in this mode due to the QCD background.
However, this is possible for $D$ because both the production rate by gluon
fusion, Eq. (36), and the decay rate into two photons, Eqs.(23,24), are
enhanced relative to the Higgs case.

For the mass range $2M_A<M_D<2M_W$, the dominant decay modes are those into
light axions so that Eq. (43) is not valid any more. Taking the number of
light axions into account, one gets $Br(D \to 2\gamma) \sim 10^{-5} \quad
\hbox{and} \sim 10^{-6}$
for models 1 and 2 respectively. The number of $D$s with $M_D \sim 100~GeV$
which can be observed in this mode in a year is 16 (which decreases to 4 for
$M_D=2M_W$) and 1 for the two models.
So in Model 1 we can use this mode to search a $D$ in the mass range whereas
in Model 2 we cannot.

We stress that all of the above mass ranges are approximate and model dependent
since the
number of PGBs, their masses and $D$ interactions are so. For a specific TC
model which is different than our generic ones, the above results will be
slightly modified.

\bigskip
\centerline{\bf 5. Conclusions}

We argue that due to the spontaneous and anomalous scale symmetry breaking
in TC there exists a pseudo--Goldstone boson of dilatations, a technidilaton,
in the low--energy spectrum of TC. The mass of the technidilaton, $D$, is
estimated to be $M_D \sim
7.5~TeV/Nr$ for a TC model with a $SU(N)_{TC}$ gauge group and $r$
technifermion
doublets. Interactions of $D$ are obtained by constructing an effective
Lagrangian which realizes both scale and chiral symmetries nonlinearly. From
this Lagrangian we see that interactions of $D$ with fermions and gauge bosons
are very similar to those of the Higgs boson with some important differences.

All $D$ interactions not induced by the trace anomaly are suppressed by a
factor of $v/F_D \sim \sqrt{0.3/N}$ with respect to the Higgs couplings.
The $D$ couplings to gauge bosons arising from the trace anomaly have two
results. First they induce $D\gamma \gamma$ and $Dgg$ interactions that have
no analog for the Higgs boson. Second they decrease the $D$ coupling to the
transverse weak gauge bosons. Even though the latter has no major
phenomenological consequences, the former plays a very important role. The
$Dgg$ coupling is not suppressed and as a result gluon fusion
is the only efficient mechanism for $D$ production. In addition, contrary to
the Higgs boson,
$D$ self--interactions get strong at high energies (i.e. $\sim TeV$) even
if $M_D<<1~TeV$ since they are proportional to energy. This might be
phenomenologically important if $M_D$ is small so that energetic $D$'s can be
produced.

The decay and production channels of the technidilaton $D$ are the same as
those for the Higgs boson except for decays into PGBs if these are
kinematically allowed. The main decay channels into $ \bar b b$,
$W^+W^-$ and $Z^0Z^0$ are suppressed by a factor of $v^2/F_D^2 \sim 0.3/N$.
The rare decays into $2\gamma$ and $2g$ are not suppressed relative to the
Higgs decays in the same channels as a result of the new couplings due to the
trace anomaly. Decays rates into PGBs are large and in general these decays
compete with decays into gauge bosons.
A heavy $D$ with $M_D>1~TeV$ is much ($\sim 15$ times for $N=6$)
narrower than a Higgs boson of the same mass which means it will be a well
defined resonance even for $M_D \sim 2~TeV$. For lower $D$ masses there are
light PGBs (since lower $D$ masses correspond to models with more technifermion
doublets) into which $D$ can decay. These decays enhance $\Gamma_{tot}$
in the mass range that they are allowed. The enhancement factor and the exact
mass ranges are model dependent. In the most exterme case, Model 2 considered
above, there are $\sim 130$ light axions and $D$ width is enhanced by about
an order of magnitude for $100~GeV<M_D<300~GeV$.

By the same token, production mechanisms of $D$ are also suppressed by the
factor $v^2/F_D^2$ which makes $D$ production very difficult compared to the
Higgs boson. Production mechanisms in $e^+e^-$ colliders, i.e. associated
production and vector boson fusion, are suppressed and therefore $e^+e^-$
colliders are not efficient for $D$ production. The cross section for
associated production is proportional to $1/s$ whereas that for vector boson
fusion increases logarithmically with $s$. We find that the latter takes over
the former at $\sqrt s=0.1~TeV+0.1M_D$. At both LEP 1 and LEP 2 the
dominant production mechanism is associated production. We find that
LEP 1 and LEP 2 can find or rule out a $D$ with $M_D \sim 10~GeV$ and
$M_D \sim 40~GeV$ respectively.
At TLC the dominant production mechanism is vector boson fusion. Unfortunately
TLC is not useful at all for producing $D$'s because of the large background
processes. The results for $e^+e^-$ colliders are model independent since
neither the production nor the decay mechanisms are model dependent.

For the SSC the situation is different since the main production mechanism
which is gluon fusion is either not suppressed (Model 2) or enhanced (Model 1).
This is a consequence of the large $D$
coupling to two gluons due to the QCD trace anomaly. $D$ phenomenology at the
SSC is model dependent since the main production mechanism which is gluon
fusion is so.
The results for $D$ observation at SSC, for the two models (the $n_T$
technifermion generation and $r$ doublet models) considered above are
as follows: For $40~GeV<M_D<2M_A \sim 100~GeV$, one can detect $D$s in the
channel $D \to 2\gamma$ in both models. This is possible only because both $D$
production due to gluon fusion and decay into two photons are enhanced as a
result of the large $D$ couplings arising from the QCD and EM trace anomalies.
The decay channel $D \to \bar b b $ is not useful because of the large QCD
background. For $2M_A<M_D<2M_W$ one can observe $D$s only in Model 1 since in
Model 2, once the light axion threshold is passed the branching ratios decrease
by an order of magnitude.

For both models the mass range $2M_W<M_D<230~GeV$ can also be
searched in the decay mode $D \to Z^0Z^0 \to l^+l^-l^+l^-$ which does not have
a large background. (One of the intermediate $Z^0$s is virtual for $2M_W<M_D<
2M_Z$.) This range cannot be extended further in Model 1 because
of the very small anomalous coupling due to small $\beta(\alpha_s)$ and in
Model 2 because
of the large number of light axions which reduce the the branching ratios by
an order of magnitude.
Summarizing the results for the SSC, in Model 1, the mass range $40~GeV<M_D
<230~GeV$ can be searched. In Model 2, on the other hand, the ranges $40~GeV<
M_D<2M_A \sim 100~GeV$ and $2M_W<M_D<230~GeV$ can be searched but the range
$2M_A<M_D<2M_W$ cannot.

\bigskip
\centerline{\bf Acknowledgements}

I would like to thank the Center for Theoretical Physics at M.I.T. where
part of this work was done. I would also like to express my gratitude to Isak
and Gulen Halyo for their support during this work.

\vfill
\eject
\centerline{\bf References }
\item{1.}L. Susskind, {\it Phys. Rev.\/} {\bf D20} (1979) 2619; S. Weinberg,
{\it Phys. Rev.\/} {\bf D19} (1979) 1277.
\item{2.}E. Halyo, {\it Mod. Phys. Lett. } {\bf A7} (1992) 201.
\item{3.}E. Halyo, {\it Mod. Phys. Lett. } {\bf A8} (1993) 275.
\item{4.}E. Farhi and L. Susskind, {\it Phys. Rep.} {\bf 74C} (1981) 277.
\item{5.}S. L. Adler, J. C. Collins and A. Duncan, {\it Phys. Rev.} {\bf D15}
(1977) 1712; J. C. Collins, A. Duncan and S. D. Joglekar, {\it Phys. Rev.}
{\bf D16} (1977) 438.
\item{6.}C. G. Callan, S. Coleman and R. Jackiw, {\it Ann. Phys.} {\bf 59}
(1970) 42.
\item{7.}R. Dashen, {\it Phys. Rev.} {\bf 183} (1969) 1245.
\item{8.}E. Witten, {\it Nucl. Phys.} {\bf B160} (1979) 57.
\item{9.}H. Georgi, {\it Weak Interactions and Modern Particle Theory\/}
(Benjamin--Cummings, 1984).
\item{10.}S. Coleman, {\it Aspects of Symmetry\/} (Cambridge University Press,
1988).
\item{11.}J. Ellis, {\it Nucl. Phys.\/} {\bf B22} (1970)478; {\it Nucl. Phys.
\/} {\bf B26} (1971) 536.
\item{12.}S. Dimopoulos and L. Susskind, {\it Nucl. Phys.} {\bf B155} (1979)
237.
\item{13.}M. Chanowitz and M. K. Gaillard, {\it Nucl. Phys.\/} {\bf B261}
(1985) 279.
\item{14.}J. F. Gunion, H. E. Haber, G. Kane and S. Dawson, {\it The Higgs
Hunter's Guide\/} (Addison-Wesley, 1990) and references therein.
\item{15.}R. Thun {\it et. al.} {\it Proceedings of the 1987 Berkeley Workshop
on Experiments, Detectors and Experimental Areas for the SSC}, ed. R. Donaldsov
and M. Gilchriese.
\item{16.}B. W. Lee, C. Quigg and G. B. Thacker, {\it Phys. Rev.\/} {\bf D16}
(1977) 1519.
\item{17.}R. Cahn and S. Dawson, {\it Phys. Lett.} {\bf B136} (1984) 196.
\item{18.}H. M. Georgi, S. L. Glashow, M. E. Mahacek and D. V. Nanopoulos
{\it Phys. Rev. Lett.} {\bf 40} (1978)692.
\item{19.}G. Altarelli and G. Parisi, {\it Nucl. Phys.} {\bf B126} (1977) 298.
\item{20.}M. E. Levi in {\it Proceedings of the 1984 SLAC Summer Institute
on Particle Physics} and references therein.
\item{21.}G. Quast, {\it Mod. Phys. Lett.} {\bf A8} (1993) 675.
\item{22.}S. Wu {\it et. al.} ``Search for Neutral Higgs at LEP 200" ,CERN
87-08
Vol. 2, June 1987.
\item{23.}J. Alcaraz, M. Felcini, M.Pieri and B.Zhou, CERN--PPE/93-28, February
1993.
\item{24.}E. Eichten, I. Hinchliffe, K. Lane and C. Quigg, {\it Rev. Mod.
Phys.}
{\bf 56} (1984) 579.
\item{25.}J. F. Gunion, G. L. Kane and J. Wudka, {\it Nucl. Phys.} {\bf B299}
(1988) 231.

\vfill
\eject

\centerline{\bf Figure and Table Captions}
Figure 1. The graph of cross section $\sigma$ as a function of the c.o.m.
energy E for
$D$ production by associated production. The long--dashed, short--dashed and
continous lines are for the cases $M_D=30,60,100~GeV$ respectively.

Figure 2. The graph of cross section $\sigma$ as a function of the c.o.m.
energy E for
$D$ production by vector boson fusion. The long--dashed, short--dashed and
continous lines are for the cases $M_D=30,60,100~GeV$ respectively.

Figure 3. The graph of $D$ production cross sections $\sigma$ at LEP 1. The
continous and dashed lines are for associated production and vector boson
fusion respectively.

Figure 4. The graph of $D$ production cross sections $\sigma$ at LEP 2. The
continous and dashed lines are for associated production and vector boson
fusion respectively.

Figure 5. The graph of $D$ production cross sections $\sigma$ at TLC. The
continous and dashed lines are for associated production and vector boson
fusion respectively.

Table 1. The cross section for $D$ production by gluon fusion for some values
of c.o.m. energy E and $M_D=100~GeV$. The cross sections $\sigma_1$ and
$\sigma_2$ refer to models 1 and 2 respectively.

Table 2. The cross section for $D$ production at the SSC for some values of
the technidilaton mass $M_D$. The cross sections $\sigma_1$ and $\sigma_2$
refer to models 1 and 2 respectively.

\end
\bye

\input tables.tex
\nopagenumbers
\magnification=1200
\baselineskip=18pt
\hbox
{\hfill
{\begintable
\ E(TeV) \ \|\ $\sigma_1(\bar p p \to D+X)$(pb) \|\ $\sigma_2(\bar p p \to
D+X)$
(pb) \crthick
    2        \|\           13          \|\       2   \nr
    10        \|\          127          \|\      21    \nr
    20        \|\          382         \|\       69    \nr
    40         \|\         850          \|\     160
 \endtable}
\hfill}
\bigskip
\parindent=0pt
\hangindent=39pt\hangafter=1

 Table 1.

 Technidilaton Phenomenology and Prospects for Production by Edi Halyo
\vfill
\eject

\end
\bye

\input tables.tex
\nopagenumbers
\magnification=1200
\baselineskip=18pt
\hbox
{\hfill
{\begintable
\ $M_D$(TeV) \ \|\ $\sigma_1(\bar p p \to D+X)$(pb) \ \|\ $\sigma_2(\bar p p
\to D+X)$(pb) \crthick
    0.1        \|\          850          \|\      160        \nr
    0.2        \|\          40           \|\       27        \nr
    0.3        \|\          0.08         \|\        4        \nr
    0.5        \|\          0.014        \|\       0.72      \nr
    1          \|\           --          \|\       0.054     \nr
    2          \|\           --          \|\       0.0018
 \endtable}
\hfill}
\bigskip
\parindent=0pt
\hangindent=39pt\hangafter=1

 Table 2.

 Technidilaton Phenomenology and Prospects for Production by Edi Halyo
\vfill
\eject

\end
\bye